\documentstyle[12pt,aasms4,tighten]{article}

\begin{document}

\font\twelvei = cmmi10 scaled\magstep1 
       \font\teni = cmmi10 \font\seveni = cmmi7
\font\mbf = cmmib10 scaled\magstep1
       \font\mbfs = cmmib10 \font\mbfss = cmmib10 scaled 833
\font\msybf = cmbsy10 scaled\magstep1
       \font\msybfs = cmbsy10 \font\msybfss = cmbsy10 scaled 833
\textfont1 = \twelvei
       \scriptfont1 = \twelvei \scriptscriptfont1 = \teni
       \def\mit{\fam1 }
\textfont9 = \mbf
       \scriptfont9 = \mbfs \scriptscriptfont9 = \mbfss
       \def\bmit{\fam9 }
\textfont10 = \msybf
       \scriptfont10 = \msybfs \scriptscriptfont10 = \msybfss
       \def\bmsy{\fam10 }

\def\etal{{\it et al.~}}
\def\eg{{\it e.g.~}}
\def\ie{{\it i.e.~}}
\def\lsim{\raise0.3ex\hbox{$<$}\kern-0.75em{\lower0.65ex\hbox{$\sim$}}} 
\def\gsim{\raise0.3ex\hbox{$>$}\kern-0.75em{\lower0.65ex\hbox{$\sim$}}} 

\title{The  Protogalactic Origin for Cosmic Magnetic Fields
   \altaffilmark{7}}

\author{Russell M. Kulsrud\altaffilmark{1,3},
         Renyue Cen\altaffilmark{1,5},
         Jeremiah P. Ostriker\altaffilmark{1,6},
    and  Dongsu Ryu\altaffilmark{2,4}}
\altaffiltext{1}
{Princeton University Observatory, Princeton, NJ 08544}
\altaffiltext{2}
{Dept.~of Astronomy \& Space Sci., Chungnam National University,
Daejeon 305-764, Korea}
\altaffiltext{3}
{e-mail: rkulsrud@astro.princeton.edu}
\altaffiltext{4}
{e-mail: ryu@sirius.chungnam.ac.kr}
\altaffiltext{5}
{e-mail: cen@astro.princeton.edu}
\altaffiltext{6}
{e-mail: jpo@astro.princeton.edu}
\altaffiltext{7}
{Submitted to the Astrophysical Journal}

\begin{abstract}

It is demonstrated that strong magnetic fields are produced from a zero initial
magnetic field during the pregalactic era, when galaxies are first forming. 
Their development proceeds in three phases. In the first phase, weak magnetic 
fields are created by the Biermann battery mechanism, acting in shocked parts
of the intergalactic medium where caustics form and intersect. In the second
phase, these weak magnetic fields are amplified to strong magnetic fields by
the Kolmogoroff turbulence endemic to gravitational structure formation of
galaxies. During this second phase, the magnetic fields reach saturation with
the turbulent power, but they are coherent only on the scale of the smallest 
eddy. In the third phase, the magnetic field strength increases to equipartition
with the turbulent energy, and the coherence length of the magnetic fields 
increases to the scale of the largest turbulent eddy, comparable to the scale 
of the entire galaxy. The resulting magnetic field represents a galactic 
magnetic field of primordial origin. No further dynamo action  is necessary, 
after the galaxy forms, to explain the origin  of magnetic fields. However,
the magnetic field may be altered by dynamo action once the galaxy  and the
galactic disk have formed.

It is first  shown by direct numerical simulations, that thermoelectric currentsassociated with the Biermann battery, build the field up from zero  to $10^{-21}$ G in the regions about to collapse into galaxies, by $z\sim3$. For weak fields, in the absence of
dissipation, the cyclotron frequency $ {\bf  \omega_{cyc} }
= e {\bf B } /m_H c $ and $ {\bf  \omega}/(1+ \chi ) $,
where $ {\bf  \omega = \nabla \times v } $ is the vorticity
and $ \chi  $ is the degree of ionization,  
satisfy the same equations,  and initial conditions
$ {\bf  \omega_{cyc} = \omega } = 0 $, so that,
 $ {\bf  \omega_{cyc}} ({\bf  r } , t) =
{\bf  \omega } ({\bf  r } , t )/(1 + \chi )$.  The vorticity  grows 
rapidly after caustics  (extreme nonlinearities) develop in the
cosmic fluid.  At this time it is shown that
turbulence has developed into Kolmogoroff turbulence.
Numerical simulations do not yet have the resolution
to demonstrate that, during the second phase,
the magnetic fields are amplified by the dynamo action
of the turbulence. Instead, an analytic
theory of the turbulent amplification of magnetic fields
is employed to explore this phase of the
 magnetic field development. 
  From this theory, it is  shown that the dynamo action of the
  protogalactic turbulence
is  able to amplify the magnetic fields, 
 during the collapse time of the protogalaxy, by such
a large factor
that  the power going into the magnetic field, 
must reach saturation with the turbulent power.
There is  as yet no analytic theory
capable of  describing the third phase.
However, preliminary turbulence calculations
currently in progress seem to confirm that
the magnetic fields proceed to equipartition
with the turbulent energy, and that the coherence
length does increase to the largest scales.
Simple physical arguments are presented that this
is the case.  Such an equipartition  field is actually too strong
to allow immediate collapse to a disk.
Possible ways around this difficulty are discussed.

\end{abstract}


\clearpage

\section{INTRODUCTION}

How do cosmic magnetic fields arise?
It has been the general belief that 
there is no natural mechanism to create a 
strong primordial
magnetic field prior to the formation of
 galaxies.  (For example, see Ruzmaikin \etal 1988.)
 For  this reason,  attention has
been concentrated on the production of 
 galactic fields  after the galactic disks  formed.
It is assumed that some weak field, 
of order $ 10^{- 17} $ G, is
initially present at the time of the
 beginning of the galactic disk,
and this weak field is amplified to its 
 present strength of a few $\mu$G
 by dynamo action of interstellar 
hydrodynamic turbulence.  (For a survey of the 
galactic dynamo theory see Ruzmaikin \etal  1988).
 But the theory for this amplification 
suffers from the problem of
 rapid amplification of small scale fields  which
inhibit the amplification of large scale fields
(Kulsrud and Anderson 1992).
In addition, the origin  of the weak ``seed field ''
is still open.  

	Because of these difficulties it is of interest to
reexamine the possibility that a strong primordial
magnetic field can be produced in the pregalactic era. 
We have done that in this paper, and we come to the
following picture of the magnetic field origin.

	 Prior
to the formation of galaxies, the universe is by no
means uniform.  Indeed, it is now known that
small relative density perturbations, $ \delta \rho/\rho $,
present at the time
of recombination, grow to finite amplitude and
form the present galaxies, clusters of galaxies,
and other structures, at a variety of epochs.  When $ \delta \rho/\rho 
\approx 0.1 $, shocks form and the resulting 
heated electrically conducting  fluid produces electric
currents  that generate magnetic fields even from 
field-free initial conditions.
	Such fields,  which are quite weak ($ \approx 10^{ -21} $
G),  would, after compression into the galactic disk,
produce the required seed field for the galactic dynamo.

	However, this is not the end of the story.  The
shocks also generate vorticity on all scales.
This  vorticity  is strong enough that its 
vortex cells turn over and generate 
Kolmogoroff turbulence. The energy spectra
of this turbulence 
has the well known power law, and
extends down to the viscous scale. (It is
true that the medium is compressible.  However, 
the shear motions in the turbulence are
themselves incompressible. This is because
 the difference in frequencies between
compressible and shear motions do not permit the 
sound waves to  interact strongly with the incompressible
shear
motions.  We can thus regard the turbulence as
incompressible and of the nature considered by 
Kolmogoroff.)  

	 The eddies
in the Kolmogoroff turbulence turn over at a rate
that is fastest for the smallest eddy.
 This smallest eddy  amplifies
the magnetic field at a rate comparable
to its turnover rate.  The 
existence of a Kolmogoroff cascade from
large eddies to small eddies  is crucial
because it enables the slowly rotating large eddies to drive
rapidly rotating small eddies,  that 
rapidly amplify the magnetic field.

	It will be shown that
conditions are always such that during the collapse
of a  protogalaxy the smallest eddy
turns over several hundred times. This  would  lead to an increase
of the magnetic field energy by a corresponding
number of powers of $ e $, the base of natural logarithms,
if saturation processes did not set in.  Instead, 
the magnetic field saturates and comes into  equipartition 
with the pregalactic turbulence. Such a saturated field 
is of sufficient strength to 
provide a primordial origin  for  galactic magnetic fields.

	If the dynamics of the magnetic field played no
role, then the magnetic field would be very chaotic on
small scales and the magnetic energy could be
concentrated on scales smaller than the smallest eddy.
  However, as the field strengthens,
the magnetic field on these  smallest scales resists  amplification,
and the  total magnetic field continues to strengthen
only on  scales comparable with that of the
smallest hydrodynamic turbulent 
eddy.  But later, when the field becomes even stronger,
the smallest hydrodynamic  eddies become suppressed
due to  the increasing drain of energy to the  magnetic field.
As a result, the spectrum of the turbulence becomes truncated
at larger scales, and the scale of the smallest hydrodynamic eddy
increases.  Eventually, only the largest
eddies survive.  During these later stages
of saturation, only the magnetic energy at the largest
scales is amplified, and the magnetic field eventually
 becomes
coherent on these largest scales.

Thus, during the pregalactic era the magnetic field goes
through three phases.  During the
first phase, thermoelectric currents generated
by shocks
increase the magnetic field strength to a 
value of order $ 10^{ -21} $ G.  This process
is known as  the Biermann  battery mechanism after
its discoverer (Biermann 1950). 
When this level of magnetic field strength is reached,
the dynamo action of the turbulence becomes faster
than the battery action,  and the second phase commences.  
During this phase,   the turbulence forms a Kolmogoroff spectrum
down to the viscous scale.  The smallest
eddy at this scale  does all the work in amplifying the magnetic field,
and continues to amplify it until saturation sets in.
	Finally, as saturation is  approached,  the third
phase takes over.  During this third  phase,  the magnetic field 
approaches equipartition with the hydrodynamic turbulence,
and becomes coherent on very large scales.

	To examine these three phases,
	we modified a standard code for the  numerical
simulation of the gravitational structure (Ryu \etal 1993).
  We first added the magnetic 
differential equation to the standard simulation
equations to study the creation and evolution 
of the magnetic field, starting from a zero initial
field. (Because the magnetic field is at first
very weak,  we neglected any magnetic forces
in the numerical simulation, so the basic 
numerical solution for velocities and pressures
was unaffected.)  We found that the hydrodynamic turbulence
does indeed create a  magnetic field, and increases
it to a field strength of about $ 10^{ -21 } $ G, 
as expected for the first phase.
However, the numerically simulated magnetic 
field did not increase beyond this point.  The reason
 is that only a small part of the Kolmogoroff turbulence 
was captured in the numerical simulation.  The
smaller eddies, which would be the most important for dynamo
amplification, were suppressed by numerical
resolution. The really important eddies
(those on the smallest scale) were on a scale
smaller than the grid size, and not seen at all.
   The only eddies
captured in the numerical simulation turn over
at too slow a rate to do more than overcome the
the damping of the field by 
numerical resistivity.  Thus,  the simulation 
has too low a resolution to correctly represent
the second phase of dynamo amplification.

	Therefore,  the remaining conclusions
concerning the enormous amplification of the magnetic
field strength expected for the second phase
can not be simulated until the numerical resolution
is improved. To properly examine this
amplification, we must employ an   analytical
method.  To do this we assume that  the Kolmogoroff spectrum  
inferred from the numerical simulation is
real. This is plausible since  the numerical evidence
for the Kolmogoroff spectrum makes this assumption 
reasonable.
 The large
scale eddies found in the simulations actually
do turn over a sufficient number of times
to produce a Kolmogoroff spectrum.
Further,  in the numerical   simulations,
where an approximately  
minus $ 5/3 $ power law was found over a limited range
in wavenumber space, this range  became larger  when 
the numerical resolution was improved.
Since the ideas of Kolmogoroff turbulence 
are now generally accepted, we felt secure in 
extrapolating  the 
spectrum to the smallest viscous scales.

	Thus, assuming that the Kolmogoroff spectrum of turbulence 
is present during phase two,
we employed a  standard theory for the dynamo  amplification
of magnetic energy by hydrodynamic turbulence in the weak
field limit (Kulsrud and Anderson 1992).
 To evaluate the total amount of
amplification we first   scaled the total turbulent  energy 
to the thermal energy. From the numerical simulations
we know that these energies are comparable. Next,  to
 find the scale of the 
smallest eddy we needed  the  plasma temperature  and density.
We normalized the temperature
to the Jean's temperature. This gives
a good approximation to the actual temperature during
the time of the collapse of the protogalaxy. Using
these results we are able to make an estimate of the
number of amplifications of the magnetic energy.
We find that this estimate  only depends only the normalization constants,
which are of order unity, and on the 
the inverse square root of the galactic mass.
For a baryonic mass in the range of a galactic mass,
this  estimate for the  number of 
e-foldings of the magnetic energy is extremely
large and demonstrates  that  the magnetic field will
probably  reach saturation with the turbulence.

	Our examination of the third phase, during
which the magnetic field is strong and drives
itself to coherence on a large scale, is still in progress
(Chandran 1996b).
However, physical arguments can be made which support
the conclusion that the magnetic field should
become coherent on  a scale comparable
with the size of the galaxy.

	Therefore, on the basis of the above
considerations and calculations, we argue that the origin 
 of galactic  fields is  primordial, i.e. predates the
galaxies in their equilibrium state. 
Even if  no magnetic field were present at the
time of recombination, a magnetic  field would be  generated
by the Biermann battery mechanism driven  by
the gravitational structure formation.
Subsequently, during gravitational collapse of the protogalaxy to the
galaxy, the residual turbulence  will drive 
dynamo action that will increase  the magnetic field strength
to saturation,  Finally, the strong  magnetic field  will smooth
itself  out on very large scales.

	The plan of the  paper is as follows:
	In section 2,  we describe the numerical simulation
which shows the initial generation of the magnetic
field, during the first phase, by the Biermann battery mechanism.
	In section 3,  we  apply the analytic theory of 
Kulsrud and Anderson (1992) to calculate the amount of dynamo
amplification during the second phase. We show that this
amplification 
is more than adequate to make   the magnetic field 
sufficiently strong that the power into it 
reaches  saturation with the hydrodynamic turbulence power.
	In section 4,  we present the physical 
arguments that imply that during the third phase
the magnetic field becomes coherent on large scales.
	In section 5,  we present our conclusions.

\section{THE NUMERICAL SIMULATION OF THE INITIAL
GROWTH OF THE MAGNETIC FIELD}

The first phase of the evolution of the protogalactic magnetic field 
can be numerically simulated by modifying
a cosmological hydrodynamic code normally used to simulate
structure formation (\eg Ryu \etal 1993).
During this phase, the 
magnetic field is too weak to be of dynamic significance
during the simulation.  Thus, we include the magnetic field
passively in the code because it does not affect the motions of the plasma. 
In the other words, we   add the evolution equation for the 
magnetic field
\begin{equation}
\frac{\partial {\bf  B}}{\partial t} =
\nabla \times ( {\bf  v } \times {\bf  B})
\end{equation}
to the code, where $ {\bf  v } $ is obtained from
the hydrodynamic part of the code without any magnetic forces.
Thus, $ {\bf  B } $ can easily be followed 
without disturbing the rest of the simulation.

The question arises concerning the initial
value  to take for the magnetic field  in the modified 
simulation.  According to equation (1),
it must be initially nonzero if $ {\bf  B } $ is to be 
nonzero in time.  A number of
proposals have been  made 
for  the generation of a small initial  magnetic field in the
early universe.  One of the more promising of these is 
 the Biermann battery mechanism, which makes use
of an extra pressure gradient term in 
Ohm's law (Biermann 1950). Employing this mechanism, we start with zero
field at the beginning of the simulation, and
include an extra term in the differential equation
for the magnetic field.  Thus,  equation (1) is
replaced by 
\begin{equation}
\frac{\partial {\bf  B}}{\partial t} =
\nabla \times ( {\bf  v } \times {\bf  B})
+ \frac{c \nabla p_e \times \nabla  n_e}{n^2_e e},
\end{equation}
where $ n_e $ is the electron density,  and $ p_e $ 
is the electron pressure. 

	 If the ionization fraction, $ \chi $,  is
taken constant in space, and the electron temperature 
is taken equal to the neutral temperature,
then $ n_e/n_B(1 + \chi ) = p_e /p $.
Here, $ n_B $  is the baryonic number density.
  From this result, equation (2) reduces to
\begin{equation}
\frac{\partial {\bf  B}}{\partial t} =
\nabla \times ( {\bf  v } \times {\bf  B})
+ \frac{ \nabla p \times \nabla  \rho }{\rho^2} \frac{cm_H}{e}
\frac{1}{1 + \chi },
\end{equation} 
where $ m_H $ is the hydrogen mass.
$\rho=n_BM/(1+\chi)$ for a hydrogen gas. 
Equation (3) only
involves the quantities which can be supplied from the hydrodynamic
part of the code, so it is easy to
incorporate the solution of  equation (3) into the
cosmological hydrodynamic code and to follow the evolution of
$ {\bf  B } $ in time as the simulation proceeds.

	For any baratropic  flow ($ p \equiv p(\rho) $),
the last source term is zero  because  $ \nabla p  $ 
is parallel to $  \nabla \rho $.
However, in general,  for a real fluid in which curved shocks
and photoheating can  occur, $ \nabla p \times \nabla \rho
\neq 0 $.

Multiplying equation (3) by $ e / m_H c $, we get the equation
for the cyclotron frequency $ {\bf  \omega_{cyc}} = e {\bf  B}/m_H c  $ 
\begin{equation}
\frac{\partial {\bf  \omega_{cyc}}}{\partial t} = 
\nabla \times ( {\bf  v } \times {\bf  \omega_{cyc} })
+ \frac{ \nabla p \times \nabla  \rho }{\rho^2}
\frac{1}{1 + \chi } + \frac{\eta c}{4 \pi} \nabla^2 {\bf  \omega_{cyc}  } .
\end{equation} 
We have added a term
to represent any resistive diffusion that may be present.

Note that the ionization fraction enters into equation (3)
through $1+\chi$, so even a very low ionization
fraction is enough to generate magnetic fields.
It is only necessary 
that there be enough electrons to carry the required current
with a drift velocity relative to the ions $v_D$ less than
their thermal velocity $v_e$, \ie $ B/ 4 \pi L = n_e e v_D/c 
<  n_e e v_e/c $ where $L$ is the scale size of variation
of the magnetic field. This condition is satisfied  by 
large margin in the numerical simulations.
( If this condition were not satisfied, an  anomalous  large
resistance would develop due to plasma instabilities, 
and the magnetic field would be  inhibited from growing.)

Equation (3) was incorporated into a numerical simulation for
the large-scale structure formation in a standard cold dark matter
(CDM) model universe with total $\Omega=1$.
The simulation was done in a periodic box with $(32h^{-1}{\rm Mpc})^3$
volume using $128^3$ cells and $64^3$ particles from $z_i=20$ to $z_f=0$.
The values of other parameters used are $\Omega_b=0.06$,
$h=1/2$, and $\sigma_8=1.05$.
For the initial condition, we adopted the standard CDM power
spectrum with $n=1$, which is modified by the transfer function
given by Bardeen \etal (1986).
Thus, the simulation is basically the same as that reported in Kang \etal
(1994), except a smaller box, and a smaller number of cells and
particles was used.

The actual equation for the magnetic field in
comoving coordinates, which we solved is
\begin{equation}
{\partial{\bf B}\over\partial t} = {1\over a}\nabla\times({\bf v}\times
{\bf B}) - 2{{\dot a}\over a}{\bf B} - {1\over B_ot_G}{m_Hc\over e}{1\over
1+\chi}{1\over a^2}{\nabla\rho\times\nabla p\over\rho^2}_,
\end{equation} 
where $B_o$ and $t_G$ are the normalization constants for the
magnetic field and time.
 This equation was solved by  using the Total Variation
Diminishing (TVD) scheme (Harten 1983). This is the same scheme
 as that employed for solution of 
 the hydrodynamic equations
(Ryu \etal 1993), so the  accuracy was comparable.
The constraint ${\bf\nabla}\cdot{\bf B}=0$ was enforced separately
in every time step.
We assumed $\chi=1$.

In Figure 1, the temporal evolution of the resulting magnetic field
is plotted.
The upper panel shows the volume-averaged (solid line) and
mass-averaged (dotted line) magnetic energy density $(B^2/8 \pi )$ as a
function of $z$.
The lower panel shows the volume-averaged (solid line) and
mass-averaged (dotted line) magnetic field strength $(B)$.
Note that $B \propto h$,  and $h=1/2$ was used.
The magnetic field strength at first grows monotonically to
the mass averaged value of order $ 10^{-21} $ G by $z\sim2-3$.
In the core of clusters it reached  $10^{-20}$ G
by $z\sim3$.
After this time, the value of the 
averaged field strength leveled off 
without further increase.
It is believed that the saturation is due to the finite numerical
resistivity inherent in the numerical scheme used to solve equation (5).

The contours of the resulting baryonic density $(\rho)$ and
magnetic field strength $(B)$ at $z=2$ are shown in Figure 2 and Figure 3.
The slice shown has a thickness of $2h^{-1}{\rm Mpc}$ (or 8 cells).
The upper panels show the whole region of $32\times32 h^{-1}{\rm Mpc}$,
while the lower panels show a  magnified region
of $10\times10 h^{-1}{\rm Mpc}$.
In Figure 2, the regions with density higher than the volume
averaged value $(0.06 {\bar\rho})$ are contoured with contour levels
$(0.06 {\bar\rho})\times10^k$ and $k=0,~0.1,~0.2,~\ldots,~2$.
Similarly in Figure 3, the regions with magnetic field strength higher
than the volume averaged value $(8\times10^{-23} {\rm G})$ are contoured
with contour levels $(8\times10^{-23} {\rm G})\times10^k$ 
and $k=0,~0.1,~0.2,~\ldots,~2$.
As expected, they are very well correlated.
Since the magnetic field was mostly generated in the accretion shocks
around the clusters, the high density core regions of the clusters have
the strongest magnetic fields.

It is interesting to compare the maximum
rms value of $ B $, or alternatively $ {\bf \omega_{cyc}} = 10^{ 4} B $,
 with the  same rms mean for the
vorticity, $ {\bf  \omega}  = {\bf \nabla} \times {\bf  v } $ 
where $ {\bf  v} $ is the fluid velocity.
The maximum rms value for $ \bf \omega $ is $ \sim10^{-16} \mbox{s}^{-1} $
around the clusters in the numerical simulation.
This is equal to the cyclotron frequency of an ion in a 
magnetic field of $ \sim10^{-20} $ G. In other words
$ e B/m_H c \approx {\bf  \omega} $ in the clusters.
This is not  surprising since
the equation for the evolution of $ - {\bf  \omega} $ is identical
to that for $ {\bf\omega_{cyc}} = e {\bf  B} /m_H c $, 
except for dissipative terms.

By taking the curl of the
equation of motion in the form 
\begin{equation}
\frac{\partial {\bf  v}}{\partial t} - {\bf v} \times ( \nabla \times 
{\bf v})  + \frac{1}{2}\nabla {\bf v}^2 = - \frac{\nabla p}{\rho}
 + \nu \nabla^2 {\bf v} 
\end{equation}
where $ \nu $ is the kinematic viscosity, one gets
\begin{equation} 
\frac{\partial {\bf \omega }}{\partial t} =
\nabla \times ( {\bf  v } \times {\bf  \omega })
- \frac{ \nabla p \times \nabla \rho }{\rho^2} + \nu \nabla^2 {\bf  \omega }.
\end{equation} 
Now we see, on  comparing equation (7) with equation (4),
that if dissipative processes are ignored (conditions well 
satisfied except during the later 
stages of the simulation), and if we assume that both $ {\bf\omega_{cyc}}$
and $ {\bf  \omega } $ are initially zero, then we should have
\begin{equation}
{\bf\omega_{cyc}} = -\frac{{\bf\omega}}{(1 + \chi )},
\end{equation} 
a remarkable result.

	 It must be appreciated that the $ \nabla p \times \nabla \rho $
term is zero until some pressure is generated,
since usually $ p $ is very small initially in the simulation.
The generation of $ p $  generally happens in shocks where 
viscosity is certainly important.
It can  be argued that the jump in 
$ {\bf\omega_{cyc}} $ and $ -{\bf  \omega}/(1 + \chi)  $ 
across a shock should be equal since, if we could treat
equation (7)  as valid through the shock,
the integral of $ \nu \nabla^2 {\bf  \omega} $ is probably
small.  Thus, $ {\bf\omega_{cyc}} $ and $ {\bf  \omega} $ 
satisfy essentially the same equation even in the shock.

A check of the above relation is presented in Figure 4.
The magnitudes of these two quantities
are displayed on a logarithmic scale.
Each point represents
the two quantities in each cell.
The magnitudes in one among eight neighboring cells
were plotted.  Here, $h=1/2$ was used again.
If the relation in equation (8) holds
exactly, all these points should lie on the line of unit slope.
The deviation for small values is presumably
due to the different dissipation rates which are not
taken into account in the derivation of this relation.
At larger values the correlation is much better,  as is
to be expected.
The rough agreement of $ {\bf\omega_{cyc}} $ and $ {\bf  \omega}/(1 + \chi) $ 
at least for larger values  tends to support  the relation in equation (8).

Eventually viscosity does become
important and $ {\bf \omega  }$  tends to saturate 
in mean square average. However, since the twisting of the
magnetic field by the $ \nabla \times ( {\bf v} \times {\bf B }) 
$  term persists, one expects that $ {\bf  B} $
will continue grow.  This fact is supported 
by Batchellor's discussion in his early paper
(Batchellor 1950).  It is thus indeed 
surprising that $ {\bf  B} $ seems to saturate
at the same time with the same amplitude as $ {\bf \omega  } $ 
does.  Is it a coincidence that numerical resistivity 
becomes important at the same time that 
viscosity does?  Since the effective viscosity is also
numerical, this coincidence is  not probably not so surprising.

\section{THE ANALYTICAL CALCULATION OF THE 
EXPONENTIAL GROWTH OF THE PROTOGALACTIC MAGNETIC FIELD
UP TO SATURATION}

Assuming the saturation of the magnetic field and $ {\bf \omega  } $
are numerical, we can argue as follows with respect  to what really 
think happens
to the magnetic field:
The value of the numerical viscosity is not much different from the 
physical viscosity, at least within a few 
orders of magnitude. But, the numerical
resistivity is larger than the physical resistivity 
by great many orders of magnitude.  Thus,
it would be natural to assume that, in the 
absence of numerical resistivity and viscosity, the magnetic field
should continue to grow,  but $ {\bf \omega  }$ should not.
This supposition is strongly
supported by Batchellor's argument(Batchellor 1950).

	To numerically  explore  the beginnings of the
second phase, in which the magnetic field strength grows
exponentially, we plan to repeat the simulation under
conditions in which the resistivity is smaller than
the viscosity.  The numerical resistivity and 
numerical viscosity are fixed by the grid size.  However,
if an additional viscous term is added
to the simulation equations, with the viscosity coefficient
clearly larger than the numerical viscosity, then
 this condition should be achieved.  Such a simulation
should enable us to confirm that the exponential growth
of the magnetic field in phase two is real. 

	The present numerical simulations have too low a resolution
to follow the behavior  of the magnetic  field into the
later phases of evolution, when dynamo
action is important. Thus,  we  turn to  the
 analytical  theory of Kulsrud and Anderson (1992)
  to estimate  the amount of the growth of the magnetic energy.
For this purpose,  we assume that during this time the turbulent 
 motions are 
represented by Kolmogoroff isotropic homogeneous
turbulence.   The normalization  constant in the Kolmogoroff spectrum
will be  determined by the numerical simulation.

Let us examine the  three-dimensional power spectra 
found in the numerical simulations.
The numerically determined 
spectra for $ (\nabla \cdot {\bf v})^2$, $P_d(k) $, and for
 $ ( \nabla \times {\bf v} )^2$, $P_c(k) $, at $ z=0 $,
  are plotted in Figure 5.  At  long wavelengths, 
the amplitude of the perturbations are
small, so that linear theory applies. In agreement with
the results of 
Peebles(1993) and others, $ P_c(k) \rightarrow 0 $ 
as $ k \rightarrow 0 $, while  $ P_d(k) $ follows the 
analytic theory expectation, $ P_d(k) \sim k^{-1}$ . 
	Thus, for very long wavenumbers,
the $ ( \nabla \times {\bf v} )^2 $ spectrum is very small,
while that of the sonic turbulence is large.  For these waves the shear
turbulence has not yet  gone nonlinear.  (Since 
$\bf \omega $ is driven by the $ \nabla p $ term as in equation (7),
it cannot really develop until shocks form.)
	
	 For
wavelengths smaller than about $ 10 h^{-1} $
Mpc, nonlinearities dominate, 
and we see that $ P_c(k)/P_d(k) \approx 1 $ .
This  spectrum is essentially the shear turbulence spectrum.
The corresponding spectrum for  sonic turbulence is
also given on this plot.

	The shear turbulence peaks at a wavenumber of
about one reciprocal Mpc . For values of $ k $
somewhat larger than this peak value,  the
spectrum  follows  a power law of 
$  k^{-5/3} $.  (This power law behavior for the 
spectrum is   expected for  Kolmogoroff 
turbulence.  This spectrum should
develop from values of $ k $ for which
the eddies have time to turn over once, since according
to Kolmogoroff this turn over generates a cascade
following the Kolmogoroff power law.  Apparently
the eddies at the peak have not yet turned over once,
since the power law spectrum does not extend all the way 
to the peak.  However, 
 at slightly larger wavenumbers, the eddies have apparently
turned over,  as 
is indicated by the fact that  the Kolmogoroff spectrum commences 
at these $ k $'s.)  The five thirds law 
does not continue to the smallest scales, but  
falls more rapidly with $ k $ as the grid scale is approached.
This cutoff is expected for the numerical simulation 
because of the large numerical viscosity.

	To demonstratethe important result that
 that the  real turbulent spectrum 
continues  to smaller  scales, one can compare the
spectrum for a finer resolution simulation (see Figure 5).
It is found that the spectrum for the finer resolution
extends  to smaller scales,  as is expected 
from the smaller effective viscosity if the
turbulence is Kolmogoroff..
This is evidence that, in the real physical situation,
there  should exist  a complete Kolmogoroff spectrum
extending down to the much smaller true  viscous scale. 

	Although the numerical results are presented 
at $ z = 0 $, the qualitative behavior should be the same
as at earlier values  of $ z $.  We apply the general Kolmogoroff theory
to earlier epochs when most galaxies were thought to have
formed,  and employ similar scalings for its amplitude. 

Let us introduce the one-dimensional spectrum
for $ v^2 $ 
\begin{equation}
{\bf v}^2 = \int I(k) dk.
\end{equation}
Then, if the turbulence is isotropic, we have
\begin{equation}
( \nabla \times {\bf v} )^2  = 4 \pi \int P(k) k^2 dk
= \int k^2 I dk,
\end{equation} 
so that $ 4 \pi P(k) = I(k) $.

We assume that the 
Kolmogoroff spectrum starts at the peak, $ k_{min} $,
and extends down to some maximum $ k , k_{max} $,
where it is truncated by physical viscosity.
That is, we write the Kolmogoroff spectrum as
\begin{equation}
I = \frac{2}{3} v_0^2 \frac{k_{min}^{2/3}}{k^{5/3}},
\end{equation}
where $ v_0^2 $ is the mean square turbulent velocity.

  From the analytic theory for the build up 
of magnetic energy given by Kulsrud and Anderson (1992) we have
\begin{equation}
\frac{d {\cal E}_M}{d t} = 2 \gamma {\cal E}_M,
\end{equation} 
where $ {\cal E}_M $ is the magnetic energy.
$ \gamma $ depends on the Kolmogoroff turbulence as
\begin{equation} 
2 \gamma \approx \int \frac{ k^2 I(k)}{\Delta \omega_k} dk,
\end{equation}
where $ \Delta \omega_k $ is the decorrelation rate at each 
$ k $ size.  We take $ \Delta \omega_k $ to be  the eddy turnover rate 
$ k \tilde{v}_k $, where $ \tilde{v}_k \approx 
\sqrt{ k I(k) } $ is the typical eddy velocity at the 
$ k $ scale.  Combining these results we find
\begin{equation} 
2 \gamma = \int  \sqrt{ k I } dk \approx {v}_0 k_{min}^{1/3}
\int^{k_{max}}_{k_{min}} \frac{dk}{k^{1/3}}
\approx \frac{3}{2}   {v}_0 k_{min}^{1/3} k_{max}^{2/3},
\end{equation}
if $ k_{max} \gg k_{min} $.  This is essentially the turnover
rate of the shortest eddy with wavenumber $ k_{max} $.

Now,
for a kinematic viscosity, $ \nu = l v_{th} $, where
$ l $ is the mean free path and $ v_{th} $  is the 
ion thermal velocity, we have
\begin{equation}
\frac{k_{max} }{k_{min}} = R^{3/4}  =
\left( \frac{v_0^2 }{k_{min} \nu }\right)^{3/4} 
=  \left( \frac{v_0^2}{v_{th} k_{min}  l}\right)^{3/4}.
\end{equation}
Therefore, 
\begin{equation}
 \gamma = \frac{3}{4} \frac{v_0^2 k_{min}^{3/2} }{v_{th}^{1/2}}
\frac{k_{min}^{1/2}}{l^{1/2}} .
\end{equation}

	A more accurate value for $ \gamma $ is derived in 
 the Appendix.  It is
\begin{equation}
\gamma=1.236 \frac{ k_0^{1/2}  v_0^{3/2}}
{l_0^{1/2}  v_i^{1/2} 
\sigma \epsilon^{1/2} }.
\end{equation} 
 In deriving this formula we set $ k_{min} = k_0  $, and assume the spectrum is
zero for $ k < k_0 $.
The mean free path, $ l_0 = 10^{ 12} T_{ev}^2/n_B $ cm,
and the thermal velocity 
$ v_i = 10^{ 6} \sqrt{ T_{ev}} $ cm/sec. 
   $ T_{ev} $ is the temperature  in electron
volts.  The constants $ \sigma $ and
$ \epsilon $ are of order unity.  They are introduced
in  the Appendix,  and are related to the model
for Kolmogoroff turbulence employed there.  The quantity 
$ 2 \gamma $ is  the exponential growth
rate of the magnetic energy in a stationary fluid,
and $ \gamma $ is the growth rate of the root
mean square field strength.

	In the expression for $ \gamma $, we expect
$ v_0 $ to be related to the thermal energy
$ 3 n_B k_B T $, where $k_B$ is the Boltzmann constant.  Let
\begin{equation}
\frac{1}{2} \rho v_0^2 = 3 \beta  n_B k_B T,
\end{equation}
where $ \beta $  is  the ratio of turbulent energy to thermal
energy. (This ratio should be of order unity
since behind a strong shock the ratio
of kinetic energy to thermal energy is 15/16.)
Thus,
\begin{equation} 
\gamma = 4.59 \frac{\beta^{3/4}}{ \sigma \epsilon^{1/2} }
 \frac{k_0^{1/2} v_i }{l_0^{1/2} } 
\end{equation}

	Let us imagine that the plasma is initially in the 
form of a sphere of radius $ R = 2 \pi / k_0 $ and collapses to form a
galaxy.  (The collapse is more likely to be two-dimensional
forming an oblate  spheroid rather than a sphere.
In this case R should refer to the shorter axis.  However,
the results on the amplification of the magnetic
field should not be very different from those
for a spherical collapse.)

	The magnetic field is also  amplified by the
collapse, as well as by the turbulence.  We  assume
that the largest turbulent eddy has the same radius as the
galaxy so $ k_0 = 2 \pi/R $.   Then, if $ B $ is the rms value of the magnetic 
field we have
\begin{equation}
\frac{d (B R^2)}{d t } = \gamma B R^2
\end{equation}
We expect the  sphere to collapse by a factor of
order 2 in the dynamic time
\begin{equation}
t_D = \frac{1}{\sqrt{ 4 \pi G n_B m_H (\rho_D/\rho_B) }}=
 \frac{8.4 \times  10^{ 14}}{\sqrt{ n_B}} \sqrt{ \frac{\rho_B}{\rho_D}}
{\rm s},
\end{equation}
where $ \rho_D/\rho_B $ is the ratio of dark matter 
density to baryonic density.  

	Thus, the amount of amplification 
during the collapse of  the sphere by a single factor of two 
is of order  $ \gamma  t_D $.
Substituting $ k_0 = 2 \pi / R $, and introducing  the 
numerical values for $ l_0 , v_i $ and $ t_D $, we get
\begin{equation}
\gamma t_D = \frac{\beta^{3/4}}{\sigma \epsilon^{1/2} }
 \frac{9.67  \times 10^{ 15}}{\sqrt{ R T_{ev} }}
\sqrt{ \frac{\rho_B}{\rho_D}}  .
\end{equation}

	 Now, at the time of collapse, we expect that the  temperature
 $ T $ should be comparable to 
the  Jean's temperature.  Therefore, we take
\begin{equation}
k_BR T_{ev} = \alpha G M m_H,
\end{equation}
where
$ M $ is the total mass
of the sphere,
dark  matter plus baryonic matter.   Let $ 
 M = 2 \times 10^{ 44} M_{11} \rho_D/\rho_B {\rm gm} $,  where
$ M_{11} $ is the baryonic mass of the galaxy in
units of $ 10^{ 11} $ solar masses.  $ \alpha $  is the
ratio of $ T $ to the Jean's temperature,  and    is
less than  unity. 
 During the collapse 
the temperature should remain isothermal because of cooling,
or at least it should remain smaller than the Jeans temperature.
For  simplicity, we take $ \alpha $ to be constant. Its actual effective
value should be  smaller than
unity because of cooling.
Thus,
\begin{equation} 
\gamma t_D = \frac{2585}{\sqrt{ M_{11}}} 
\frac{\beta^{3/4}}{\alpha^{1/2}\sigma \epsilon^{1/2} } 
 \left( \frac{\rho_B}{\rho_D} \right) .
\end{equation}

	To form a more precise estimate of the 
 factor by which $ B R^2 $ grows,  we need the integral 
\begin{equation}
 A = \int \gamma dt.
\end{equation}
Consider as an example a uniform-density,  isolated 
sphere of radius $ R $.  Let the initial radius be
$ R_0 $, the initial value of $ \gamma $ be $ \gamma_0 $,
and the initial value of $ t_D $ be $ t_{D0} $. 
	  Let $ R = x R_0 $.
 During the  collapse of the sphere, 
the pressure of the baryonic matter is negligible so
that the  equation for the time dependence of $ x $ is
\begin{equation}
\frac{d^2 x}{d t^2}  = - \frac{4 \pi }{3} G \rho_0 
\frac{1}{x^3},
\end{equation} 
where $  \rho_0 $ is the initial density. 

	Integrating  equation (26) we have 
\begin{equation}
\frac{1}{2} \left( \frac{d x}{d t} \right)^2 = \frac{4 \pi G \rho_0}{3}
\frac{1}{x} + const.
\end{equation}
For simplicity we neglect this constant, so that
\begin{equation}
\frac{dx}{dt} = \sqrt{\frac{2}{3}} \frac{1}{ t_{D0} x^{1/2} }.
\end{equation}
 
	From equation (24) we have that as time varies 
$ \gamma \sim t_D^{-1} \sim \rho^{1/2} \sim x^{-3/2} $.
Using this scaling, we can now integrate $ \gamma $ in time
\begin{equation}
A= \int \gamma dt = \int \gamma \frac{d x}{dx /dt} =
\gamma_0 t_{D0} \sqrt{ \frac{3}{2}}  \int \frac{dx}{x}.
\end{equation}
Substituting for $ \gamma_0 t_{D0} $ from equation (24),
we have
\begin{equation}
A= \int \gamma dt = \frac{3166}{\sqrt{ M_{11}} }\frac{\rho_B}{\rho_D}
 \frac{\beta^{3/4}}{\sigma \epsilon^{1/2} \alpha^{1/2}}
  \ln \frac{1}{x} .
\end{equation}
	Taking $ \rho_B/\rho_D  = 1/20 $ and $ x= 1/2 $,  we have
\begin{equation}
A= \int \gamma dt =\frac{109.7}{\sqrt{ M_{11}}}
   \frac{\beta^{3/4}}{\sigma \epsilon^{1/2} \alpha^{1/2}}. 
\end{equation} 
	 The number of powers of ten by which $ R^2 B $ increases during
the dynamic time, $ A_{10} $,  is
\begin{equation} 
A_{10} =  \log_{10} e \int \gamma dt  
 = \frac{47.6}{\sqrt{ M_{11}}} 
\frac{\beta^{3/4}}{\alpha^{1/2} \epsilon^{1/2} \sigma }.
\end{equation}

\section{THE EVOLUTION OF THE MAGNETIC FIELD AFTER 
SATURATION}

	An increase in the rms magnetic field strength 
by this 
large  number of powers of ten,
clearly indicates  that  $ B $ should grow to reach  saturation with
the turbulent power.  Just how
the saturation proceeds is somewhat
complicated, and deserves further analysis.
However, from the results of Kulsrud and Anderson 
(1992),  we can make some remarks as to how we think that the 
saturation should proceed.

	First, before saturation occurs, 
during the period when  the magnetic energy grows exponentially, 
the wavenumber at which the magnetic energy is concentrated
propagates to very small wavelengths. The peak
wavenumber $ k_{peak} $  of the magnetic 
 spectrum, $ M_B(k) $, increases exponentially,
as $ k_{peak}  \approx e^{ \gamma t/2} $. 
The magnetic spectrum itself increases as $ k^{3/2} $
 up to $ k_{peak}$, and then 
falls off rapidly at larger wavenumbers. At any fixed $ k $, 
 the magnetic spectrum 
grows in time  as $ e^{3 \gamma t/4} $.
The total integrated magnetic energy, $ {\cal E}_M
 = \int M_B(k) $, thus, increases as
as $ e^{2 \gamma t} $ as it should.

	Second, the magnetic energy at scale
$ k $ has an alfven 
 frequency $ \omega_A(k) = k \sqrt{ 2 {\cal E}_M / \rho }$ (Chandran 1996a).
As $ {\cal E}_M  $ increases and  $ k_{peak} $ increases,
there comes a time when the frequency $ \omega_A(k_{peak} ) $ 
exceeds the eddy turnover rate $ \gamma $.  When this happens, 
 the
eddy frequency  become mismatched with the alfven frequency, 
and amplification of magnetic energy at these scales
is no longer possible.
	This means that  for scales such that $ k > k_{crit} $, where
\begin{equation} 
\omega_A(k_{crit}) = k_{crit} \sqrt{ 2 {\cal E}_M /\rho } = \gamma, 
\end{equation}
the magnetic  energy no longer grows exponentially.  
It is seen that as $ {\cal E}_M $ increases, $ k_{crit} $ decreases.
It can be shown  that,  by the time that  $ k_{crit} $ becomes as small as 
$ k_{max} $, the wavenumber of the smallest
turbulent eddy,  the magnetic energy is concentrated at 
$ k_{max} $.

	Third, at this latter time $ 2 \gamma {\cal E}_M $ is comparable to the
turbulent power  $( \approx k\rho v_0^3/2) $.  After this time
 the magnetic energy
stops growing exponentially and starts growing linearly in time.
This transition happens because  the drain on the turbulent power
by the magnetic field is stronger than the  drain by
viscous dissipation.  Hence, the hydrodynamic spectrum
is truncated by the drain of energy into the magnetic
energy, rather than by viscous dissipation. 
As a consequence  $ k_{max} $ decreases
at just such a rate that $ \gamma $, which depends on $ k_{max}$ 
as $ k_{max}^{2/3} $, 
 decreases so as to keep $ 2 \gamma {\cal E}_M $ always  equal
to the turbulent power. At the time when  saturation 
commences, the 
cutoff wavenumber $ k_{crit} = k_{max} $. 
Subsequently,  the wavenumber range 
in which the magnetic energy is predominant continues to
be at $ k_{max} $,  and this wavenumber becomes
 smaller and smaller. In the end,
when essentially all the turbulent energy has been 
converted to magnetic energy, the magnetic energy is all concentrated
near  $ k_{max} \approx  k_{min} $.  

	To determine if this last phase is reached, we must
ask  whether, during phase two  when exponential growth
occurs, the magnetic energy at $ k_{max} $ 
 reaches the value $ {\cal E}_{sat} $, given 
\begin{equation}
\frac{3}{4} \gamma  {\cal E}_{sat} = \frac{1}{2} 
k_{min}\rho v_0^3.
\end{equation}
This is the value of magnetic energy at which  the 
 turbulent power  into the magnetic field 
 is comparable with the total turbulent power.
(At this time, the turbulent power at smaller scales has
stopped growing.   The rate of 
 growth of magnetic energy is
 $ (3/4) \gamma $,  rather than $ 2 \gamma $.)
If this condition is satisfied, then we expect that all the turbulent energy
will convert to magnetic energy.  Thus, $ {\cal E}_{sat} $   is a
critical value for $ {\cal E}_M $.  At this magnitude  and beyond,
all the magnetic energy is in a band less than $ k_{max} $.
The energy in this limited range has grown   only by the factor
 $ e^{ 3 \gamma t/4 } $.  Thus, during  collapse of the 
galaxy, the important number is the number  of e-foldings of
the magnetic energy in the range $ k_0 $ to $ k_{max} $.
For total saturation it 
must be large enough for $ {\cal E}_M $ to reach this critical value.
Since number of e-folds of  the  energy in this limited range
is $ (3/4) \int \gamma dt $,  the number of powers
of ten by which the effective value of $ B R^2 $ increases
during the collapse 
is reduced
from that of equation (32),  by the factor $ 3/8 $,  that is to
\begin{equation}
A'_{10} = \frac{3}{8} A_{10}
 = \frac{ 17.87}{\sqrt{ M_{11}}} \frac{\beta^{3/4} }
{\alpha^{1/2} \epsilon^{1/2} \sigma } .
\end{equation}
If this number is large enough to amplify the
magnetic field up to the saturated value such that
\begin{equation}
\frac{3}{4} \gamma {\cal E}_M = \frac{1}{2}
 k_{min}\rho v_0^3,
\end{equation}
then the remaining turbulence energy will be  converted
to magnetic energy at a linear rate, and the field 
will become  coherent on the scale of the largest turbulent 
eddy,  $ k_{min}^{-1} $.  On the other hand,
if this number is too small to reach the saturated
value of the magnetic field before the turbulence 
damps, then the resulting magnetic field strength is smaller, and
the magnetic field is much more chaotic.  

	For the determination of the critical number,
$ A'_{10} $, we need values for $ \alpha, \beta, \sigma,$
 and $ \epsilon $.  For a lower limit on $ A'_{10} $ 
we may take $ \alpha = 1 $, since we expect the temperature
at the commencement of collapse to certainly be less
than the Jean's temperature $ T_J $.  A realistic
estimate of $ \sigma, $ where $ \sigma \Delta \omega_k $
is the breadth frequency spectrum of the turbulence at
fixed $ k $, is about $ 2 $.  (The definitions of these
parameters are  given just before equation (A4).)
A value of $ \epsilon $ of $ 1 $ seems plausible where
the parameter $ \epsilon $ is related to the rate of  
turbulent transfer (see equation (A14)).  
To find $ \beta $ we make use of the numerical simulation.
$ \beta $ is the ratio of turbulent kinetic  energy to thermal
energy. If we assume that the turbulent energy is one
half of the total kinetic energy, then we can take
the ratio of total kinetic energy density to thermal
energy to be $ 2 \beta $.  This ratio depends on
the ratio of the density to the mean density, 
 $ \rho / <\rho> $. Its value taken from the numerical 
simulation,   is plotted
in Figure 6.  The appropriate  value for  
$ \rho/<\rho> $ at the time of collapse is
$ 5 $ (Peebles 1993).   
  From  Figure 6 we find from this value
for the relative density 
that $ \beta \approx 5 $.  Thus,  substituting 
$ \alpha = 1$, $\beta = 5$, $\sigma =2 $, and $ \epsilon = 1 $,
into equation (35) we find that the number powers
of ten by which $ B $ increases is
\begin{equation}
A'_{10} = 30.
\end{equation}
This number is certainly sufficiently large  to show that $ B $  increase 
from the value $ 10^{ -21} $ G obtained 
from
the Biermann battery mechanism, to the critical  value
of $ B $ 
 for saturation with the turbulent power. 
(For the value of $ \gamma $ found from equation (24)
with our choice of parameters,  this critical field strength for
 $ B $ is about $ 10^{ -7} $ G.
If   this critical value for the  magnetic field strength 
is reached, then we expect equipartition
with the total turbulence energy to result.  This
final value of $ B $ is about $ 10^{ -5} $ G.)

\section{CONCLUSION}

	In conclusion, we have described a mechanism
for generating the observed cosmic magnetic fields
  that proceeds in three  phases.

	In the first phase, the Biermann battery mechanism, 
driven by cosmic turbulence originating in shocks
and caustic formation, creates a magnetic field from a zero
initial field.
This mechanism is found to produce  a magnetic field
whose strength is about $ 10^{-21} $ G.  
In the second phase,
Kolmogoroff turbulence amplifies this field  by a factor 
sufficient to reach saturation with the turbulent power.
(Neglecting saturation the factor of increase would
be of order $ 10^{ 30} $.) 
	In the third phase, the magnetic field reaches equipartition
with the turbulent energy, yielding a field strength of order
$ 10^{ -5 } $ G.  During this phase it becomes
coherent on the scale of 
the largest turbulent eddy.  This scale is comparable
with the scale of the entire galaxy.

	The final value of the magnetic field, which we predict  to be 
in equipartition with the turbulent  energy,
 is very strong.  In fact, it is too
strong to allow interstellar matter to 
flatten into a disk once it separates from 
the dark matter halo.  The likely state of affairs at this time 
is that the  sphere of interstellar plasma
cannot immediately collapse to a disk.   Probably, what happens is that
star formation occurs removing a substantial
fraction of the interstellar plasma from the
magnetic field.  The remnant of the plasma
is then too light to hold the magnetic field,
and the  bulk of the  remaining plasma is
expelled back into the intergalactic or
intra cluster medium carrying the field with it.
The remaining field can then be compressed
by the interstellar medium, and  enriched by matter
thrown off by the
evolved halo  stars.  Such a process should
leave some observational trace, which might
be used to check this theory of the protogalactic origin
of the cosmic field.

	It should be remarked that the dynamo action that
we have examined has no $ \alpha $ effect such
as is usually postulated for mean field dynamo theory.
These theories are kinematic and ignore any
effect of magnetic forces.  Thus, to produce
a coherent field in the kinematic limit,  some coherence in the turbulence
is necessary.  However, when the field becomes
very strong, it generates its own coherence,
since it is impossible to put a large energy
into small scales.  This is  because the magnetic tension
would  become  too strong if the magnetic field strength 
is increased on small scales.  The magnetic field is perfectly
capable of unwinding itself, and producing
its own coherence, when it is sufficiently
strong.  The necessity for an $ \alpha $ effect
is purely a weak field result, only applicable
in the kinematic limit.

	It is interesting that the triple phase 
process of creation of cosmic magnetic fields was 
suggested by Biermann himself many years ago(Biermann 1950).

\acknowledgments
The authors are grateful for a helpful discussion
with  Achterburg,  Ben Chandran, Steven Cowley, Ravi Sudan,
Martin  Rees, and   Ethan Vishniac.
The work of Kulsrud was  supported by the  National Science
foundation under Grant AST 91-21847
and by  NASA's astrophysical program
under  grant  NAGW 2419.
The work by Ryu was supported in part by the Basic Science
Research Institute Program, Korean Ministry of Education 1995,
Project No.~BSRI-95-5408. The work by Cen and Ostriker was
supported by the National Science Foundation Grant AST-9318185.

\clearpage

\begin{center}
{\bf FIGURE CAPTIONS}
\end{center}

\begin{description}

\item[Fig. 1] Temporal evolution of the magnetic field.
The upper panel shows the volume-averaged (solid line) and
mass-averaged (dotted line) magnetic energy density $(B^2/8\pi)$ as a
function of $z$.
The lower panel shows the volume-averaged (solid line) and
mass-averaged (dotted line) magnetic field strength $(B)$.

\item[Fig. 2] Density contours of a slice with a thickness of
$2h^{-1}{\rm Mpc}$ (or 8 cells) at $z=2$.
The contour lines with density higher than $0.06 {\bar\rho}$ are shown
with levels $0.06\times10^k$ and $k=0,~0.1,~0.2,~\ldots,~2$.
The upper panel shows the whole region of $32\times32 h^{-1}{\rm Mpc}$,
while the lower panel shows the magnified region of
$10\times10 h^{-1}{\rm Mpc}$.

\item[Fig. 3] Magnetic field strength contours of a slice with a
thickness of $2h^{-1}{\rm Mpc}$ (or 8 cells) at $z=2$.
The contour lines with magnetic field strength higher than
$8\times10^{-23} {\rm G}$ are shown with levels
$8~10^{-23}\times10^k$ and $k=0,~0.1,~0.2,~\ldots,~2$.
The upper panel shows the whole region of $32\times32 h^{-1}{\rm Mpc}$,
while the lower panel shows the magnified region of
$10\times10 h^{-1}{\rm Mpc}$.

\item[Fig.4] Magnitude of $ {\bf \omega}/(1+\chi) $ plotted against
that of $ {\bf \omega_{cyc}} $ on a logarithmic scale. 
Each point represents the values in each cell.
One among eight neighboring cells were plotted.
The predicted relation is the forty-five degree
straight line. The  correlation is quite good for the larger
values.

\item[Fig. 5]  Three-dimensional power spectra
for $ (\nabla \cdot {\bf v})^2$ (solid curves),
\begin{equation}
\langle(\nabla \cdot {\bf  v } )^2 \rangle = \int P_d(k)  d^3 {\bf  k},
\end{equation} 
and for $ ( \nabla \times {\bf v} )^2$ (dashed curves),
\begin{equation}
\langle(\nabla \times {\bf  v } )^2 \rangle = \int P_c(k)  d^3 {\bf  k},
\end{equation} 
produced by numerical simulations.
The thick (solid and dashed) curves are from 
a power law ($P_k=k^{-1}$) hydrodynamic 
simulation with boxsize $L=80h^{-1}$Mpc
and $512^3$ fluid elements, and
the thin (solid and dashed) curves are from 
a power law ($P_k=k^{-1}$) hydrodynamic 
simulation with boxsize $L=90h^{-1}$Mpc
and $288^3$ fluid elements.
The $L=80h^{-1}$Mpc simulation
has a resolution half as that of the $L=90h^{-1}$Mpc simulation.
Comparison of the results from the two simulations
indicates that the steep decline at the high $k$ end of the
power spectra is due to limited numerical resolution.
Also shown is a straight dotted line which is parallel
to the Kolmogoroff spectrum of logarithmic slope $-5/3$.

\item[Fig. 6]  Ratio of the local kinetic  energy
to the thermal energy density in a numerical simulation, 
as a function of the local relative density
$ \rho / <\rho> $.  $ <\rho> $ is the mean density.
If the turbulence kinetic energy density is one half
of the total kinetic energy density, then this ratio is equal to 
$ 2 \beta $,  where $ \beta $   is the normalizing
parameter introduced  in the text as the ratio
of turbulent energy density to thermal energy density.

\end{description}

\clearpage

\appendix
\section{A More Precise Value for the Growth Rate $ \gamma $}

In this appendix, we evaluate the 
growth rate of the rms magnetic field for the
Kolmogoroff spectrum of the turbulent
energy per unit mass per unit wavenumber,
$ I(k) $. $ I(k) $  is the one-dimensional spectrum 
defined in Eq. (9).  We assume that $ I $ is zero for $ k< k_0 $,
and in the inertial range $ k \ll k_{max} $ where $ k_{max} $
 is the cutoff wavenumber, $ I(k) $ is given by
\begin{equation} 
I(k) = \frac{2}{3} v_0^2 {\frac{k_0^{2/3}}{k^{5/3}}}_.
\end{equation}
Here, $ v_0^2  $ is the mean square  value of the 
turbulent velocity.
$I(k)$ suffers a viscous cutoff at $ k_{max} $.
We make use of the formalism of Kulsrud and Anderson(1992).
  
In Kulsrud and Anderson,
$ v_0^2/ 2 $ is given in terms of the $ k , \omega $ spectrum,
 $ J(k, \omega ) $,
the energy per  wavenumber mode,
per unit angular frequency, per unit mass in
a box normalization of size $ L $, 
\begin{equation} 
\frac{v_0^2}{2} = \left(\frac{L}{2\pi}\right)^3 
\int d^3 k d \omega J(k, \omega ),
\end{equation}
while $ \gamma $ is expressed in term of the
quantity
\begin{equation}
U(k)= \left(\frac{L}{2\pi}\right)^3 2 \pi J(k, 0),
\end{equation}
which involves the zero frequency harmonic of the
turbulence.

We assume the frequency spectrum at each
$ k $ has a width $\sigma ( \Delta \omega )_k $,
where
\begin{equation} 
(\Delta \omega)_k =  k \tilde{v}_k,
\end{equation}
and $\sigma$ is a constant of order unity.
Here, 
\begin{equation}
\tilde{v}_k^2 = \int_{k/2}^{2k} I(k) dk
\end{equation}
is the contribution from
a range about $ k $ extending from $ k/2 $ to $ 2k $.
\begin{equation}
\tilde{v}^2_k = \frac{3}{2}  k I \left[2^{2/3} - (1/2)^{2/3}\right]
= 1.436   k I,
\end{equation}
so that
\begin{equation} 
( \Delta \omega )_k = 1.198 k^{3/2} I^{1/2}.
\end{equation}

Thus,  we have
\begin{equation} 
\int J(k,\omega ) d \omega =  J(k,0) \sigma(\Delta \omega)_k.
\end{equation}
Combining the above equation with Eqs. (A2) and (A3) and the definition of
$I(k)$, we get
\begin{equation}
U(k) = \frac{I(k)}{4 k^2  \sigma (\Delta \omega)_k}.
\end{equation}
Substituting Eq. (A7) in this equation, we get
\begin{equation}
U(k) = 0.2086 {\frac{I(k)^{1/2}}{\sigma k^{7/2}}}_.
\end{equation}

Now,  in the kinematic
limit, according to Kulsrud and Anderson(1992), 
the total magnetic energy $ {\cal E}_M $ grows  at the
rate 
\begin{equation}
\frac{d {\cal E}_M }{d t} = 2 \gamma {\cal E}_M,
\end{equation}
where according to their Eq. (2.44)
\begin{equation} 
\gamma = \frac{1}{3} \int_{k_0}^{k_{max}} k^2 U d^3 k
=  \frac{4 \pi }{3} \int_{k_0}^{k_{max}} k^4 U d k.
\end{equation}
Substituting Eq. (A10) for $ U $,  we get
\begin{equation}
\gamma = 0.8738  \int_{k_0}^{k_{max}} 
\frac{(kI)^{1/2} }{\sigma  }  d k.
\end{equation}

The Kolmogoroff spectrum 
$ I(k) $  is given by Eq. (A1)  for $ k  $ not
too near $ k_{max} $.  In order to find a more complete expression for
$ I $ near $ k_{max}$,  we carry out a dimensional analysis
of the Kolmogoroff spectrum.  

We assume that, due to mode coupling, all the energy
from a band $ k/2 $ to $ k $ is moved to the band
$ k $ to $ 2k $ in a time $ 1/\epsilon (\Delta \omega)_k $.
Here, $\epsilon$ is a constant of order unity,
Thus, the flux of energy passing through $ k $,  $ \Phi $,  is
\begin{equation}
\Phi = \frac{1}{2} \epsilon (\Delta \omega )_k \int_{k/2}^k I dk 
= \epsilon (\Delta \omega)_k \frac{3}{4} (2^{2/3}- 1) k I 
= 0.4406  \epsilon (\Delta \omega )_k k I.
\end{equation}
In the evaluation of the constant  we have employed 
the power law approximation for the inertial range of  $ I(k) $, Eq. (A1).

Now, according to Braginski, energy is 
dissipated by viscosity at the
rate
\begin{equation}
\eta \left[ 2 \left(\frac{\partial v_z}{\partial z}\right)^2
+ \left(\frac{\partial v_x}{\partial x}\right)^2
+\left(\frac{\partial v_y}{\partial y}\right)^2 \right],
\end{equation} 
where $ z $ is the prevailing direction of the
rms magnetic field.  The above result is correct as
long as the gyroradius is small.  Averaging this result
over all directions, we get the energy dissipated per unit 
$ k $ is
\begin{equation}
\frac{\eta k v^2}{5 }=  {\frac{\eta k^2 I }{5 }}_.
\end{equation}
Thus, equating the derivative of the flux $ \Phi $ to
this viscous damping rate, we get the approximate
equation for $ I $ 
\begin{equation}
0.4408 \frac{\partial}{\partial k} \left[ \epsilon (\Delta \omega)_k k I
\right]
= - {\frac{\eta k^2 I}{5}}_.
\end{equation}

The solution to this equation that agrees with Eq. (A1)  is
\begin{equation}
I(k)= \frac{2}{3} \frac{ v_0^2 k_0^{2/3} }{k^{5/3}} \left[
1 - \left(\frac{k}{k_{max} }\right)^{4/3} \right]_,
\end{equation}
where
\begin{equation}
k_{max} = 5.033{\frac{v_0^{3/4}  k_0^{1/4}}{\eta^{3/4}}}_.
\end{equation}

Substituting Eq. (A18) into Eq. (A13), we get
{\begin{equation}
\gamma = \frac{4 \pi }{18 \sigma } k_0^{1/3} k_{max}^{2/3} v_0.
\end{equation}
  Further substituting Eq. (A19) for $ k_{max} $, we get
\begin{equation}
\gamma = 2.050 {\frac{v_0^{3/2} k_0^{1/2}}{\sigma \epsilon^{1/2} \eta^{1/2}}}_.
\end{equation}

As given by Braginski, the ion viscosity can be expressed as
\begin{equation}
\eta =  2.75 v_i l_0,
\end{equation}
where $v_i$ is the ion thermal velocity and and $l_o$ is the ion mean
free path given by
\begin{equation}
v_i = 10^{ 6} T_{ev}~~\mbox{cm/s} 
\end{equation}
\begin{equation}
l_0 = 10^{ 12} T_{ev}^2/n_B~~\mbox{cm}.
\end{equation}
Here, $ T_{ev} $ is the ion temperature expressed in
electron volts and $ n_B $ is the ion density.
Substituting Eq. (A22) into (A21), we get
\begin{equation}
\gamma = 1.236 {\frac{v_0^{3/2} k_0^{1/2}}
{\sigma \epsilon^{1/2}v_i^{1/2} l_0^{1/2}}}_,
\end{equation}
which is Eq. (17).


\clearpage

\begin{thebibliography}{}

\bibitem[]{}
Bardeen, J. M., Bond, J. R., Kaiser, N., \& Szalay, A. S.,
1986, \apj, 304, 15.

\bibitem[]{}
Batchellor, G. K., 1950, Proc. Roy. Soc. London., A201, 405.

\bibitem[]{}
Biermann, L.,  1950, Z. Naturforsch, 5a, 65.

\bibitem[]{}
Chandran, B., 1996a, submitted to \apj.

\bibitem[]{}
Chandran, B., 1996b, private communication.

\bibitem[]{}
Harten, A., 1983, J. Comp. Phys., 49, 357.

\bibitem[]{}
Kang, H., Cen, R., Ostriker, J. P., \& Ryu, D., 1994, \apj, 428, 1.

\bibitem[]{}
Kulsrud, R. M., \& Anderson S. W., 1992, \apj, 396, 606.

\bibitem[]{}
Peebles, P.J.E., 1993, ``Principles of Physical Cosmology'',
Princeton University Press, Princeton.

\bibitem[]{}
Ruzmaikin, A.A., Shukurov, A.M. \& Sokoloff, D.D.,
1988, ``Magnetic Fields in Galaxies'', Astrophysics and
Space Science Library, Kluwer Academic Publishers,
Dordrecht.

\bibitem[]{}
Ryu, D., Ostriker, J. P., Kang, H., \& Cen R., 1993, \apj, 414, 1.



\end{thebibliography}
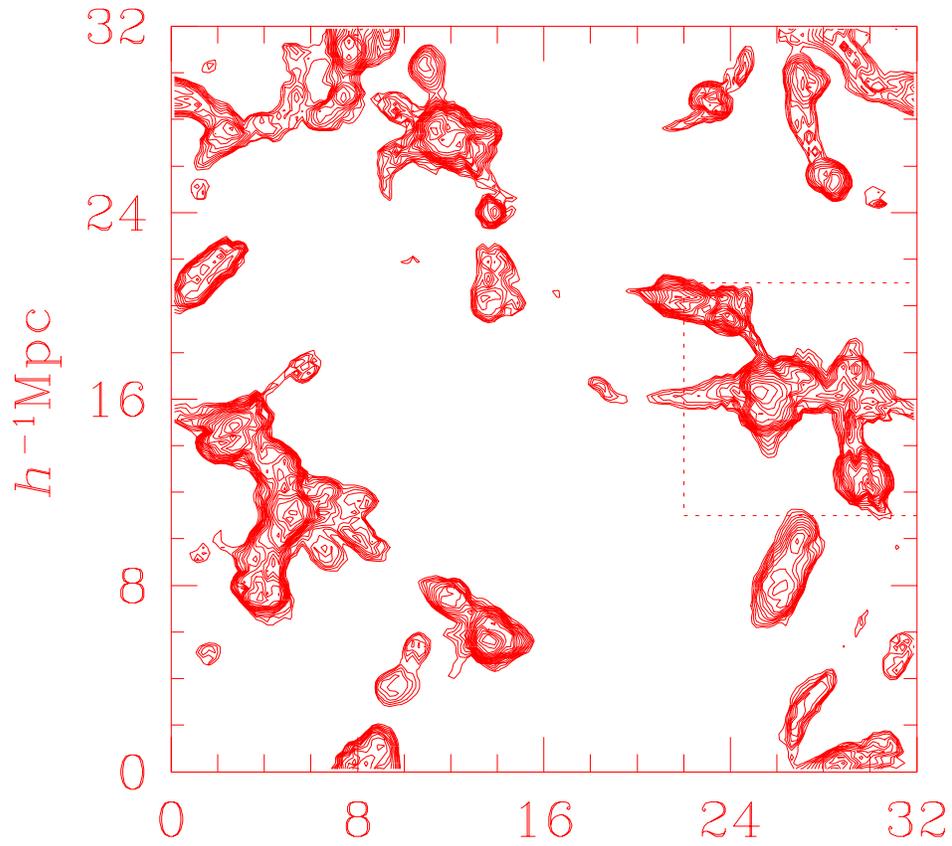
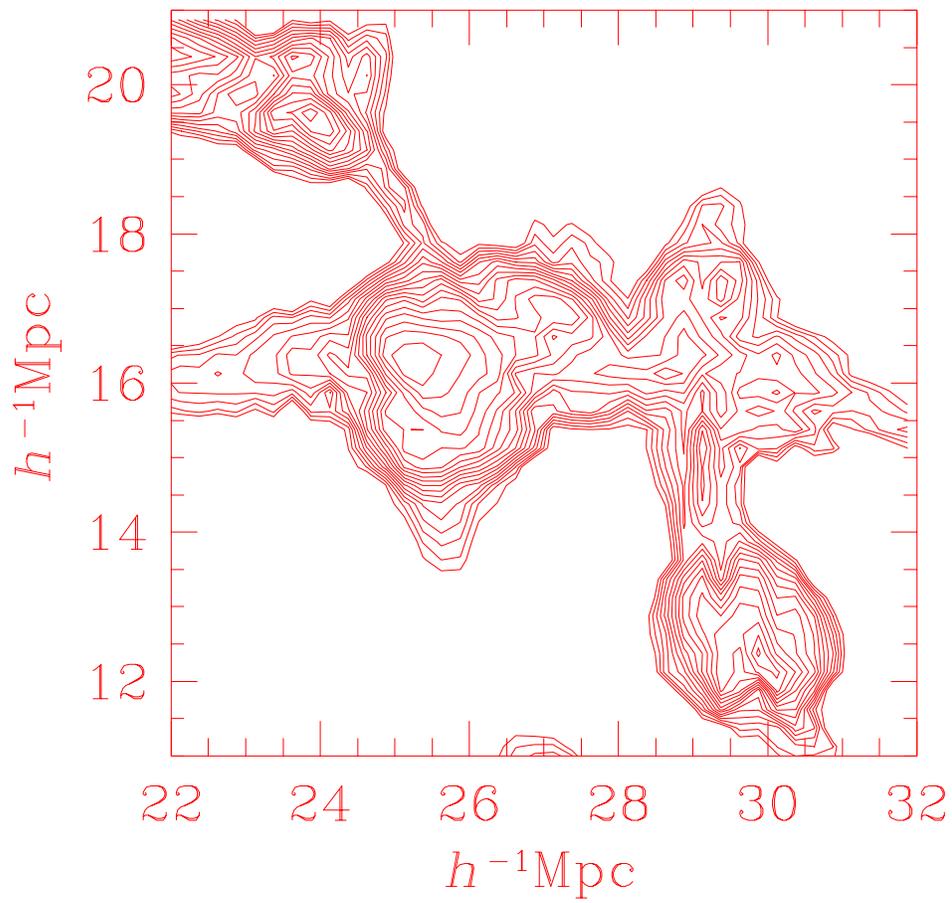
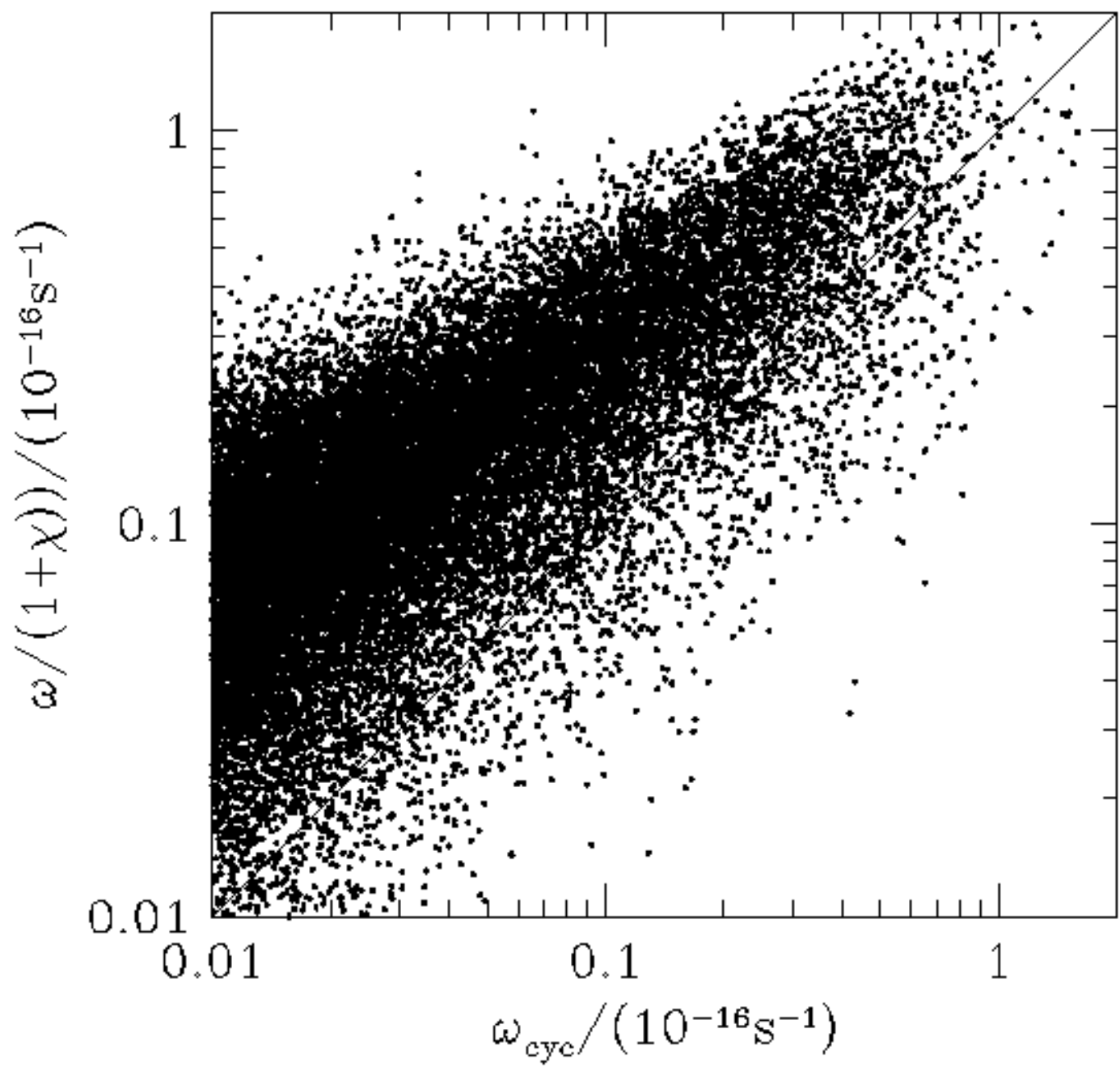
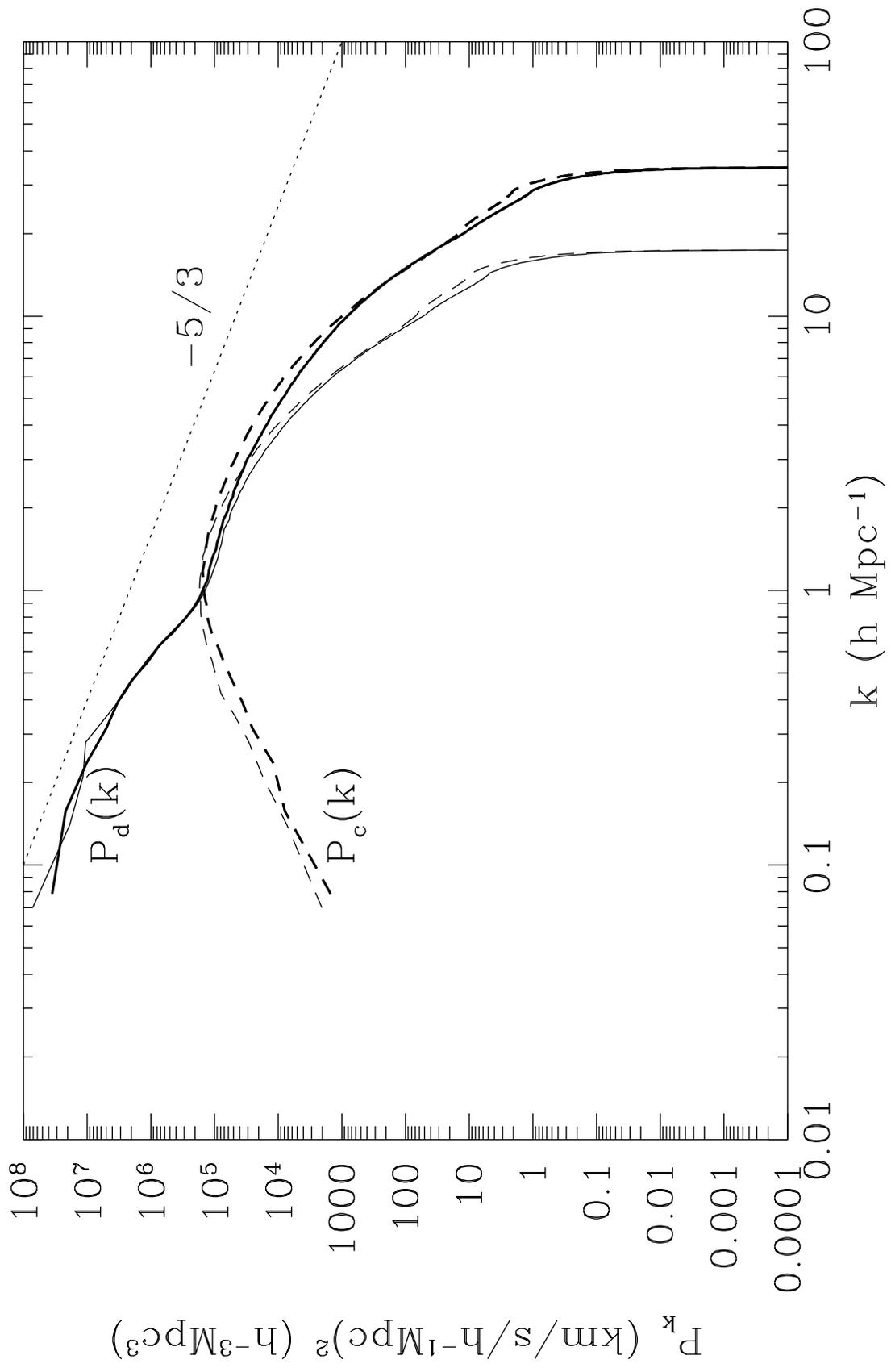
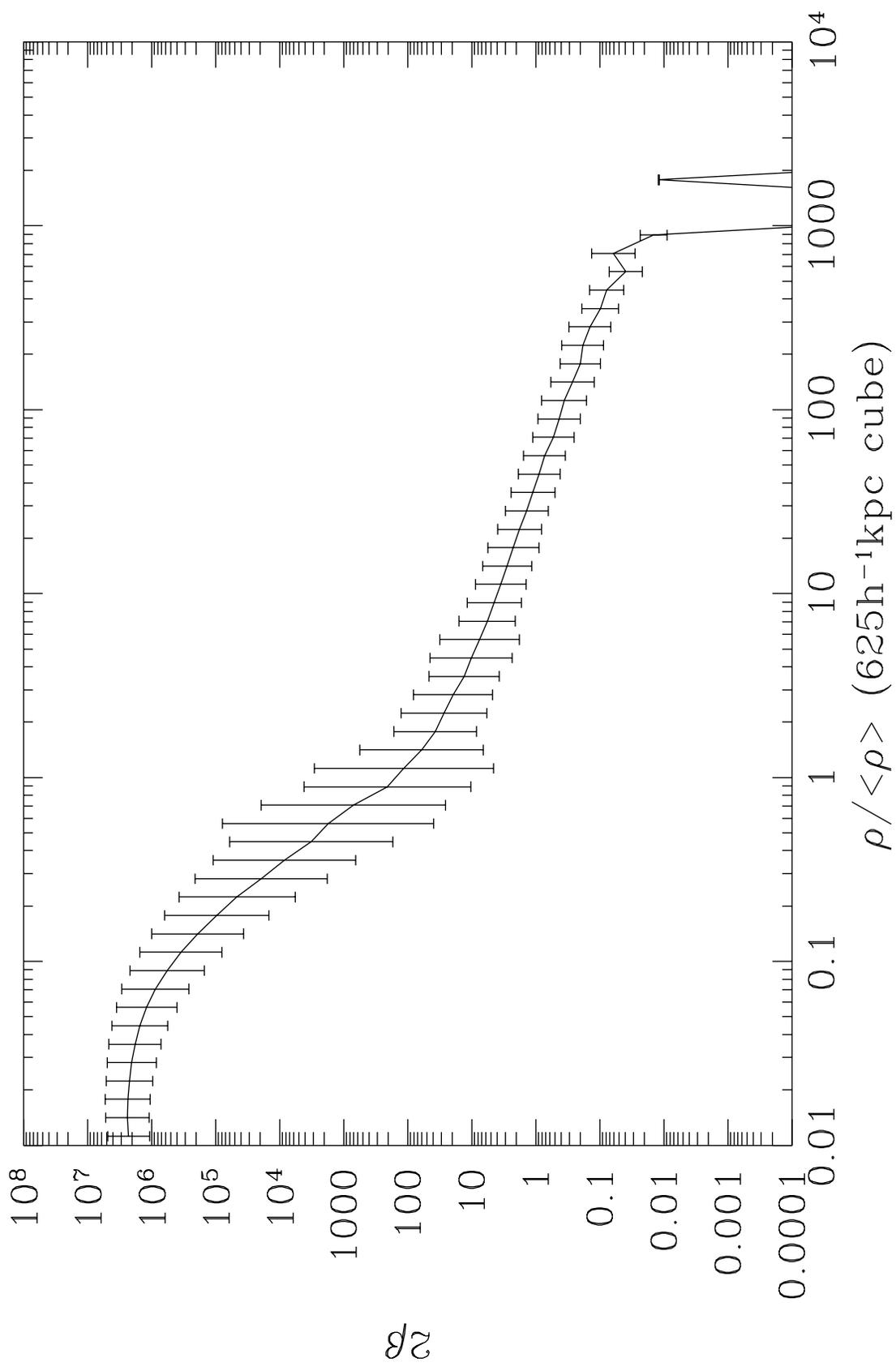
\end{document}